\documentclass[10pt]{iopart}
\usepackage{graphicx}
\usepackage{iopams}

\begin{document}

\title[$\pi/K/p$ production from p+p, d+Au and Au+Au collisions at PHENIX]{$\pi/K/p$ production and Cronin 
effect from p+p, d+Au and Au+Au collisions at $\sqrt{s_{NN}}~=~$200~GeV from the PHENIX experiment.}

\author{Felix Matathias\dag for the PHENIX\footnote[3]{For the full PHENIX Collaboration author list 
and acknowledgments , see Appendix ``Collaborations'' of this volume.} collaboration}
\address{\dag Department of Physics and Astronomy, 
\mbox{State University of New York at Stony Brook},
 Stony Brook, NY, 11794}

\begin{abstract}

We present results on identified particle production in p+p, d+Au and Au+Au collisions  at  $\sqrt{s_{NN}} =
200$~GeV at mid-rapidity measured by the PHENIX experiment. The centrality and flavor dependence of the Cronin effect in
d+Au collisions is measured. The Cronin effect for the protons in d+Au is larger than that for the pions, but
not large enough to account for the ``anomalous'' proton to pion ratio in central Au+Au collisions.

\end{abstract}



% \maketitle

%%%%%%%%%%%%%%%%%%%%%%%%%%%%%%%%%%%%%%%%%%%%%%%%%%
% \section{Introduction}
%%%%%%%%%%%%%%%%%%%%%%%%%%%%%%%%%%%%%%%%%%%%%%%%%%

In the search for the Quark Gluon Plasma at the Relativistic Heavy Ion Collider 
it was discovered that hadron production at  
high transverse momentum ($p_T \ge 2$\,GeV/$c$) is suppressed in central 
Au+Au collisions~\cite{Adcox:2001jp} compared to 
nucleon-nucleon collisions. Furthermore, the yields of $p$ and $\bar{p}$ near 2\,GeV/$c$ in central 
events~\cite{Adler:2003kg,Adler:2003cb} are comparable to the yield of pions ($p/\pi\sim 1$), in contrast
to the proton-to-pion ratios of $\sim 0.1 - 0.3$ in p+p~\cite{Alper:1975jm}.
Novel mechanisms of particle production in the environment of 
hot and dense nuclear matter were proposed~\cite{Vitev:2001zn, Hwa:2002tu, Greco:2003xt, Fries:2003vb} to 
explain this phenomenon. 
The determination of the properties of this environment is not straightforward though and requires 
control over the interplay between many competing nuclear effects~\cite{Vitev:2002pf} that 
influence the particle spectra. 
These include shadowing, enhanced production at moderate $p_T$ \mbox{(Cronin effect~\cite{Cronin:zm})},  and energy loss.
It is therefore of paramount importance to perform a control experiment to
quantify these competing mechanisms and distinguish effects 
of the surrounding cold nuclear matter (initial state effects) 
from those originating from highly excited nuclear matter (final state effects).

Consequently, RHIC collided d+Au at $\sqrt{s_{NN}}~=~$200~GeV, and also
provided p+p collisions for a baseline measurement.
PHENIX measured identified particle spectra, yields, ratios, and nuclear
modification factors from these data.
The d+Au experiment, besides serving  as a comparison measurement for relativistic heavy ion 
collisions, acquires a significance of its own. 
The centrality dependence of particle production in d+Au 
collisions can provide detailed information on the microscopic origin of the Cronin effect and possibly 
provide insights on the baryon enhancement.
Also, protons exhibit a larger Cronin effect than pions and kaons at lower energies~\cite{Straub:xd}
and at RHIC~\cite{Adams:2003qm} while no theoretical calculations can account for the species dependence of the effect.

\begin{figure}
\center
\includegraphics[width=9 cm]{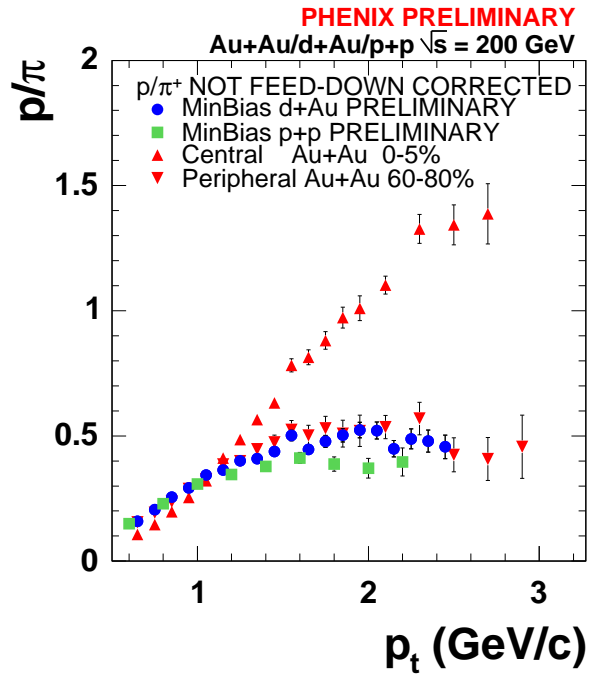}
\caption{Inclusive proton to pion ratio in 3 colliding systems: p+p, d+Au and Au+Au. The protons are not feed-down corrected.}
\label{fig:figure1}
\end{figure}

%%%%%%%%%%%%%%%%%%%%%%%%%%%%%%%%%%%%%%%%%
% \section{Data analysis}
%%%%%%%%%%%%%%%%%%%%%%%%%%%%%%%%%%%%%%%%%

The data were collected in 
RUN02 for Au+Au and p+p collisions at $\sqrt{s_{NN}} = 200$~GeV; in 
RUN03, d+Au collisions at the same energy were studied. A detailed presentation of the Au+Au analysis can
be found in~\cite{Adler:2003cb}. In the following we discuss the analysis of the p+p and d+Au data. 
Minimum bias events with vertex position along the beam axis within $|z| <$~30 cm were 
triggered by the Beam-Beam Counters (BBC) located at  $|\eta|$~=~3.0-3.9. 
The minimum bias trigger accepts (88 $\pm$ 4\%) of all inelastic d+Au collisions and (51.6 $\pm$ 9.8\%) of
all p+p inelastic collisions that satisfy the vertex condition. 
A total of 1.4 $\times$ 10$^7$ d+Au events and 1.35 $\times$ 10$^7$ p+p events were analyzed.

Charged particles are tracked using the central arm spectrometers.
The spectrometer on the east side of the PHENIX detector (east arm) 
contains the following subsystems used in this analysis: drift chamber (DC), pad chamber (PC1) 
and time-of-flight (TOF). Identified charged particles are measured using a portion of the  
east arm spectrometer covering pseudo-rapidity $\Delta\eta = 0.7$ and
$\Delta\phi=\pi/8$ in azimuthal angle. Particle identification is based on particle mass calculated from the 
momentum measured by the DC and the velocity obtained from the time-of-flight  
and the path length along the trajectory.

%%%%%%%%%%%%%%%%%%%%%%%%%%%%%%%%%%%%%%%%%%%%
% \subsection{Centrality selection in d+Au}
%%%%%%%%%%%%%%%%%%%%%%%%%%%%%%%%%%%%%%%%%%%%

Collision centrality is selected in d+Au using the south (Au-going side)
BBC (BBCS). We assume that the BBCS signal is proportional to the number of participating 
nucleons ($N^{Au}_{part}$) in the Au nucleus, and that the hits in the BBCS are uncorrelated 
to each other.
Using a Glauber simulation of d+Au collisions the distribution of $N^{Au}_{part}$ is generated for different
impact parameters.
For a given centrality the charge distribution of the BBCS is 
described by a negative binomial distribution.
Parameters are assumed to be proportional to $N^{Au}_{part}$,
and fitted to reproduce the data, yielding
a BBCS charge response function.
 We define 4 centrality classes in d+Au 
collisions with $N_{coll}$=$15.0\pm1.0$, $10.4\pm0.7$, $6.9\pm0.6$ and $3.2\pm0.3$.

%%%%%%%%%%%%%%%%%%%%%%%%%%%%%%%%%%%%%%%%%%%%%%
% \section{Results}
%%%%%%%%%%%%%%%%%%%%%%%%%%%%%%%%%%%%%%%%%%%%%%
\begin{figure}
\center
\includegraphics[width=10 cm]{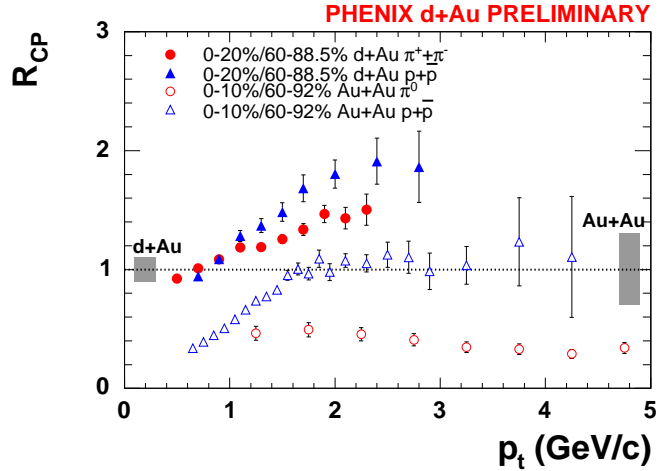}
\caption{Nuclear modification factor $R_{CP}$ for pions and protons in d+Au and Au+Au collisions.}
\label{fig:figure2}
\end{figure}

The proton to pion ratio from p+p, d+Au and Au+Au is shown in Figure~\ref{fig:figure1}. 
The protons are not feed-down corrected in this ratio since this correction was not available for the
d+Au and p+p data. For feed-down corrected $p$ and $\bar{p}$ spectra in Au+Au see~\cite{Adler:2003cb}. 
The $p/\pi$ ratio in d+Au is very similar to the peripheral Au+Au while the p+p ratio is somewhat lower.
It is evident that the high value of the $p/\pi$ ratio in central Au+Au collisions can only be attributed
to the hot and dense nuclear matter created in this class of collisions while the baseline measurement is much lower.

Figure~\ref{fig:figure2} shows the central to peripheral 
ratio in d+Au and Au+Au for $N_{coll}$ scaled $p_{T}$ spectra of $(\bar{p} + p)/2$ and pions. 
We define the central to peripheral ratio $R_{\rm CP}$ as: 
\begin{equation}
 R_{\rm CP} =\frac{{\rm Yield^{central }}/\langle N_{coll}^{central} \rangle}
                  {{\rm Yield^{peripheral}}/\langle N_{coll}^{peripheral} \rangle}.
\label{eq:rcp}
\end{equation}
Although the protons have a larger Cronin effect than the pions in d+Au, on the order of 20-25\%, this phenomenon can not
account for the approximately factor of 3 enhancement of the protons with respect to the pions in the Au+Au case.
Therefore the enhancement of the protons relative to the pions in central Au+Au is not just due to the species-dependent Cronin effect. 
The baryon production mechanism must depend on the produced medium; even if it is already present in d+Au, the effect is 
necessarily much smaller \cite{Hwa:2004zd}.

\begin{figure}
\center
\includegraphics[width=10 cm]{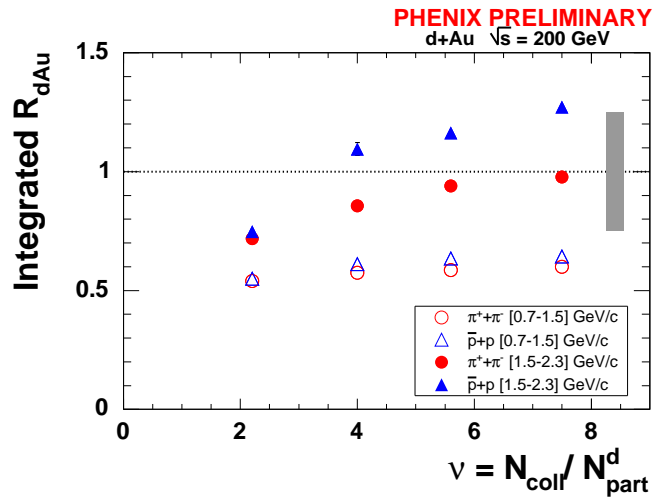}
\caption{Integrated $R_{dAu}$ for pions and protons.}
\label{fig:figure3}
\end{figure}

Figure~\ref{fig:figure3} shows $R_{dAu}$, the nuclear modification factor using the 
p+p spectrum as a reference, for pions
and protons in two momentum bins. The modification factors are plotted
as a function of the number of collisions suffered by the deuteron
participant nucleons. The lower momentum bin is in the $p_T$ region
where $R_{dAu}$ is less than 1.0, and the dependence on the number of
collisions is minimal. In the higher $p_T$ bin, $R_{dAu}$  shows a steady increase with
the number of collisions. The rate of increase is larger for baryons than for mesons.

Traditional explanations of the Cronin effect involve multiple soft scattering 
of incoming partons that lead  to an enhancement at intermediate $p_{T}$. There are various
theoretical models~\cite{Accardi:2002ik} of the multiple scattering, predicting
somewhat different dependence upon the number of scattering centers.
Recently, Hwa and collaborators~\cite{Hwa:2004zd} have shown 
an alternative explanation of the Cronin effect, attributed to 
the recombination of shower quarks with those from the medium,
even in d+Au  collisions.  

%%%%%%%%%%%%%%%%%%%%%%%%%%%%%%%%%%%%%%%%%%
%\section{Conclusion}
%%%%%%%%%%%%%%%%%%%%%%%%%%%%%%%%%%%%%%%%%%
In summary, we have presented the centrality and species dependence of identified particle spectra in
all 3 collision systems studied at RHIC to date.
The proton to pion ratio in d+Au is similar to the peripheral Au+Au while the corresponding ratio in p+p is somewhat lower.
The nuclear modification factor in d+Au for protons shows a larger Cronin effect than that for pions but not large enough
to account for the abundance of protons in central Au+Au collisions.

\end{document}